# Perpendicular magnetic anisotropy in Pt/Co-based full Heusler alloy/MgO thin films structures

M.S. Gabor[1a)], M. Nasui[1], A. Timar-Gabor[2]

[1]*Center for Superconductivity, Spintronics and Surface Science, Physics and Chemistry Department, Technical University of Cluj-Napoca, Str. Memorandumului, 400114 Cluj-Napoca, Romania*

[2] *Interdisciplinary Research Institute on Bio-Nano-Sciences, Babeş-Bolyai University, Str. Treboniu Laurean, 400271, Cluj-Napoca, Romania*

*Abstract*

Perpendicular magnetic anisotropy (PMA) in ultrathin magnetic structures is a key ingredient for the development of electrically controlled spintronic devices. Due to their relatively large spin-polarization, high Curie temperature and low Gilbert damping the Co-based full Heusler alloys are of special importance from a scientific and applications point of view. Here, we study the mechanisms responsible for the PMA in Pt/Co-based full Heusler alloy/MgO thin films structures. We show that the ultrathin Heusler films exhibit strong PMA even in the absence of magnetic annealing. By means of ferromagnetic resonance experiments, we demonstrate that the effective magnetization shows a two-regime behavior depending on the thickness of the Heusler layers. Using Auger spectroscopy measurements, we evidence interdiffusion at the underlayer/Heusler interface and the formation of an interfacial CoFe-rich layer which causes the two-regime behavior. In the case of the ultrathin films, the interfacial CoFe-rich layer promotes the strong PMA through the electronic hybridization of the metal alloy and oxygen orbitals across the ferromagnet/MgO interface. In addition, the interfacial CoFe-rich layer it is also generating an increase of the Gilbert damping for the ultrathin films beyond the spin-pumping effect. Our results illustrate that the strong PMA is not an intrinsic property of the Heusler/MgO interface but it is actively influenced by the interdiffusion, which can be tuned by a proper choice of the underlayer material, as we show for the case of the Pt, Ta and Cr underlayers.

[a)] mihai.gabor@phys.utcluj.ro

*Introduction*

Ultrathin films structures showing perpendicular magnetic anisotropy (PMA) are under intensive research for the development of electrically controlled spintronic devices. Particularly, current induced spin–orbit torques (SOTs) in heavy-metal/ferromagnet (FM) heterostructures showing PMA are used to trigger the magnetization switching[1,2]. Besides, the antisymmetric interfacial Dzyaloshinskii-Moriya[3,4] interaction (iDMI) in similar PMA architectures, if strong enough, can lead to the formation of special chiral structures like skyrmions[5] which are drivable by electrical currents[6]. In the case of the spin transfer torque magnetic random-access memories (STT-MRAMs), considered as a potential replacement for the semiconductor-based ones, the use of strong PMA materials is required for increased thermal stability[7], while high spin polarization and low Gilbert damping is needed to obtain large magnetoresistive ratios and efficient current induced STT switching[8,9].

Cobalt-based full Heusler alloys are a special class of ferromagnetic materials that attract an increased scientific interest, since their theoretical prediction of half-metallicity[10]. These compounds are described by the formula $Co_2YZ$, where Y is a transition metal, or a mixture of two transition metals, and Z is a main group, or a mixture of two main group *sp* elements. Large magnetoresistive ratios are experimentally demonstrated in certain $Co_2YZ$ based in-plane[11-18] and out-of-plane[19,20] magnetized magnetic tunnel junctions (MTJs) and relatively low Gilbert damping parameters were determined for some compounds[21-28]. Furthermore, PMA was evidenced for $Co_2FeAl/MgO$[19,29-33], $Co_2FeAl_{0.5}Si_{0.5}/MgO$[34-37], $Co_2FeSi/MgO$[38,39] or $Co_2Fe_xMn_{1-x}Si/MgO$[40,41] structures with different non-magnetic underlayers. In some of the cases, an annealing stage was necessary to induce PMA, while for the other the perpendicular magnetization was achieved even in the as-deposited state. The origin of the strong PMA in this type of structures is still under debate. It could be related to both the oxidation at the Heusler/MgO interface[42,43] and to the spin-orbit interaction effects at the heavy-metal underlayer/Heusler interface[44,45]. Moreover, it was recently pointed out that in the case of $Co_2FeAl/MgO$ the diffusion of Al towards the MgO layer during annealing plays an important role for the stabilization of the PMA[46,47]. The precise knowledge and control of the mechanisms responsible for PMA is essential in order to be able to develop viable spintronic applications. Therefore, in this paper, we study the underlying physics governing the PMA for $Co_2FeAl$, $Co_2FeAl_{0.5}Si_{0.5}$, $Co_2FeSi$ and $Co_2Fe_{0.5}Mn_{0.5}Si$ Heusler alloy thin films sandwiched between Pt and MgO layers. We show that below a certain critical thickness all the Heusler films show strong PMA even in the absence of magnetic annealing. Additionally, using ferromagnetic resonance experiments, we demonstrate that, depending on the thickness of the Heusler layers, the effective perpendicular magnetic anisotropy shows a two-regime behavior. After excluding other possible mechanisms, we evidence using Auger spectroscopy measurements, that the diffusion of the lighter elements towards the Pt underlayer and the formation of an interfacial CoFe-rich layer causes the two-regime behavior. In the case of the ultrathin films, this interfacial CoFe-rich layer promotes the strong PMA through the hybridization of the [Co,Fe] $3d_{z^2}$ and O $2p_z$ orbitals at the interface and is also responsible for the increased Gilbert damping. Our study reveals that the strong PMA is not intrinsic to the Heusler/MgO interface. It is strongly influenced by the interdiffusion and can be adjusted by a proper choice of the underlayer material, as we show for the case of the Pt, Ta and Cr underlayers.

*Experimental*

All the samples studied here were grown at room temperature on thermally oxidized silicon substrates in a magnetron sputtering system having a base pressure lower than $2\times10^{-8}$ Torr. The main samples have the following structure: Si/SiO$_2$//Ta (3 nm)/Pt (4 nm)/ FM (0.8-10 nm)/MgO (1 nm)/Ta (3 nm), where FM stands for Co$_2$FeAl (CFA), Co$_2$FeAl$_{0.5}$Si$_{0.5}$ (CFAS), Co$_2$FeSi (CFS), Co$_2$Fe$_{0.5}$Mn$_{0.5}$Si (CFMS) or CoFeB (CFB), depending on the sample. Additional samples were grown, and their structure will be discussed later in the text. The metallic layers were deposited by dc sputtering under an argon pressure of 1 mTorr, while the MgO layer was grown by rf sputtering under an argon pressure of 10 mTorr. The Heusler alloys thin films were sputtered from stoichiometric targets. The 3 nm thick Ta buffer layer was grown directly on the substrate to minimize the roughness and to facilitate the (111) texturing of the upper Pt layer. The 1 nm thick MgO layer was deposited to induce perpendicular magnetic anisotropy on the Heusler thin film[7]. An additional 3 nm thick Ta capping layer was sputtered to protect the samples from oxidation due to air exposure. The structure of the samples was characterized by x-ray diffraction (XRD) using a four-circle diffractometer. The static magnetic properties have been investigated using a Vibrating Sample Magnetometer (VSM), while the dynamic magnetic properties by using a TE(011) cavity Ferromagnetic Resonance (FMR) setup working in X-band (9.79 GHz). Auger spectra have been recorded in derivative mode, using a cylindrical mirror analyzer spectrometer working at an electron beam energy of 3keV. Depth profile analysis have been performed by successive recording of the Auger spectra and Ar ion sputter-etching of the surface of the samples by using a relatively low ion energy of 600 eV.

*Results and discussions*

Figure 1(a) shows *2θ/ω* x-ray diffraction patterns recorded for four representative Pt (4 nm)/ $Co_2YZ$ (10 nm)/ MgO (1 nm) samples. Irrespective of the Heusler composition, the patterns show the (111) and (222) peaks belonging to the Pt layer, the (022) peak arising from the Heusler films and the (001) peak of the Si substrate. This indicates that the Pt layer has a (111) out-of-plane texture, while the Heusler films are (011) out-of-plane textured. Laue oscillations are observable around the (111) Pt reflection which confirms the good crystalline quality for the Pt films[48]. Moreover, ϕ-scan measurements (not shown here) indicate that both the Pt underlayer and the Heusler films have no in-plane texturing but show an in-plane isotropic distribution of the crystallites. No peak belonging to the Ta capping layer was observed, indicating that the film is in an amorphous or nanocrystalline state.

The static magnetic properties of our films were characterized by VSM measurements. Figure 2 shows hysteresis loops measured with the magnetic field applied perpendicular to the plane of the samples, for representative Heusler films thicknesses. In order to remove the substrate diamagnetic contribution, we fitted the large field data with a linear function and extracted the linear slope from the raw data. Regardless of their composition, all the Heusler films show a similar behavior. Above a critical spin-reorientation transition thickness the samples show in-plane magnetic anisotropy. This is indicated by the shape of the hysteresis loops in Fig. 2 (a)–(d), which is typical for a hard axis of magnetization, showing a continuous rotation of the magnetization up to saturation. Below this critical thickness, the samples show PMA, which is attested by the square shaped hysteresis loops in Fig. 2 (e)–(h). We also determined the saturation magnetization ($M_S$) and the effective thicknesses of the ferromagnetic layers using hysteresis loop measurements and the procedure described in [31]. The effective thicknesses of the ferromagnetic layers are used throughout the paper and the $M_S$ is found to be 790 ± 70 emu/cm³, 660 ± 50 emu/cm³, 935 ± 75 emu/cm³ and 895 ± 75 emu/cm³ for CFS, CFMS, CFAS and CFA samples, respectively.

In order to get more insights on the magnetic anisotropy properties of our films, we have performed FMR measurements with the magnetic field applied at different $θ_H$ angles (defined in the inset of Fig. 4) with respect to the normal direction of the layers. Figure 3 shows typical FMR spectra for various field angles recorded for a 2.4 nm thick Pt/CFAS sample. We define the resonance field $H_R$ as the intersection of the spectrum with the base line, and the linewidth $H_{PP}$ as the distance between the positive and negative peaks of the spectrum. Figure 4 shows the $θ_H$ dependence of the $H_R$ and of the linewidth $H_{PP}$ for the 2.4 nm thick Pt/CFAS sample. In order to extract the relevant FMR parameters, we analyzed the $\theta_H$ dependence of the FMR spectrum using a model in which the total energy per unit volume is given by

$$E = -M_S H \cos(\theta_H - \theta_M) + 2\pi M_S^2 \cos^2\theta_M - K_\perp \cos^2\theta_M, \tag{1}$$

where the first term is the Zeeman energy, the second term is the demagnetizing energy, and the last term is the magnetic anisotropy energy. The $M_S$ is the saturation magnetization, $\theta_H$ and $\theta_M$ are the field and magnetization angles defined in the inset of Fig. 4, and the $K_\perp$ is the effective perpendicular magnetic anisotropy constant. From eq. 1 and the Landau-Lifshitz-Gilbert equation, one can derive the resonance condition as[49]

$$\left(\frac{\omega}{\gamma}\right)^2 = H_1 \times H_2, \tag{2}$$

where $\omega$ is the angular frequency of the microwave, $\gamma$ is the gyromagnetic ratio, given by $\gamma = g\mu_B\hbar$ where $g$ is the Landé g-factor, $\mu_B$ is the Bohr magneton and $\hbar$ is the reduced Planck constant, and with $H_1$ and $H_2$ given by

$$H_1 = H_R \cos(\theta_H - \theta_M) - 4\pi M_{\text{eff}} \cos^2\theta_M, \tag{3}$$

$$H_2 = H_R \cos(\theta_H - \theta_M) - 4\pi M_{\text{eff}} \cos 2\theta_M, \tag{4}$$

where $4\pi M_{\text{eff}}$ is the effective magnetization defined as $4\pi M_{\text{eff}} = 4\pi M_S - 2K_\perp/M_S$ and $H_R$ is the resonance field. For each value of $\theta_H$, the $\theta_M$ at resonance is calculated from the energy minimum condition $\partial E/\partial \theta_M = 0$. Hence, the $H_R$ dependence on $\theta_H$ can be fitted by Eq. (2)-(4) using $4\pi M_{\text{eff}}$ and $g$ as adjustable parameters. A typical fit curve is shown in Fig. 4(a).

Figure 5 shows the $g$ factor dependence on the thickness of the Heusler layers for samples with different Heusler layer compositions. Depending on the thickness, two regimes are discernable. For relatively large thicknesses, above 2.5-3 nm, the $g$ factor shows rather constant values between 2.07 and 2.11, depending on the type of the Heusler layer. For lower thicknesses, $g$ shows a monotonous decrease, regardless of the Heusler layer composition. This is an interface effect and it is usually attributed to the fact that at the interfaces, due to the symmetry breaking, the orbital motion is no longer entirely quenched and will contribute to the gyromagnetic ratio[50,51]. Another possibility, which cannot be excluded in our case, is the reduction of the $g$ factor due to intermixing between the ferromagnetic Heusler layer and non-magnetic materials at the interfaces[50].

Figure 6 shows the effective magnetization $4\pi M_{\text{eff}}$ dependence on the inverse thickness of the ferromagnetic layer for samples with different compositions. It is to be mentioned that the $4\pi M_{\text{eff}}$ was determined from FMR experiments only for samples with in-plane magnetic anisotropy (positive $4\pi M_{\text{eff}}$). In the case of ultrathin samples showing perpendicular magnetic anisotropy (negative $4\pi M_{\text{eff}}$), due to the strong linewidth enhancement, it was not possible to obtain reliable resonance curves. Therefore, in this case the $4\pi M_{\text{eff}}$ was estimated from VSM measurements. Generally, it is considered that the effective perpendicular magnetic anisotropy constant $K_\perp$ can be written as the sum of a volume ($K_V$), which includes magneto-crystalline and strain related anisotropies, and a surface ($K_S$) contribution: $K_\perp = K_V + K_S/t$, where $t$ is the thickness of the ferromagnetic layer. Thus, the effective magnetization can be written as

$$4\pi M_{\text{eff}} = \left(4\pi M_S - \frac{2K_V}{M_S}\right) - \frac{2K_S}{M_S}\frac{1}{t}. \qquad (5)$$

The above relation implies a linear dependence of the effective magnetization on the inverse thickness of the ferromagnetic layer. However, as shown in Fig.6 the Heusler samples do not show a single linear dependence for the entire thickness range, but two regimes above and below a certain critical thickness. Using the $M_S$ values determined from VSM measurements and by fitting the experimental data in the large thickness regime to eq. (5), we extract a surface anisotropy constant $K_S$ for the CFA and CFAS of 0.24 ± 0.03 erg/cm² and 0.22 ± 0.02 erg/cm² and a volume contribution $K_V$ of (1.27 ± 0.69)×10⁶ erg/cm³ and (1.51 ± 0.7)×10⁶ erg/cm³, respectively. In the case of the CFMS and CFS, the $K_S$ was negligible small within the error bars and the $K_V$ was found to be (0.44 ± 0.38)×10⁶ erg/cm³ and (0.51 ± 0.4)×10⁶ erg/cm³, respectively. Using the as extracted values of the anisotropy constants, we can calculate, for example, in the case of the CFAS samples a spin-reorientation transition thickness of around 0.55 nm. This is clearly not in agreement with the experimental data, as seen from Fig. 2 and 6, already a 1 nm thick CFAS film shows strong PMA and it is spontaneous perpendicularly magnetized. This is a consequence of the fact that the 1 nm thick CFAS film falls within the second anisotropy regime below the critical thickness. The occurrence of this second anisotropy regime with larger effective perpendicular magnetic anisotropy can have several explanations. For such thin films one must always consider the possible influences of the surface roughness. If the roughness is relatively large, an in-plane demagnetization field will develop at the edges of the terraces which will reduce the shape anisotropy and favor perpendicular magnetization. This is equivalent to the emergence of an additional dipolar surface anisotropy contribution[52]. The roughness is a parameter which is not easily quantifiable experimentally in such thin multilayer structures. However, it is reasonable to expect to be comparable for similar heterostructures in which the Heusler alloy film is replaced with a CFB layer. Atomic force microscopy topography images (not shown here) recorded for heterostructure with CFB and CFAS layers are featureless and show a similar RMS roughness. As such, if the low thickness anisotropy regime is due to the roughness it must be observable also in the case of CFB samples. However, this is not the case, as shown in Fig. 6, the CFB samples show a single linear behavior for the whole range of thickness. Fitting the data to eq. (5), allowed us to extract for CFB samples a surface anisotropy contribution $K_S$ of $0.79 \pm 0.04$ erg/cm$^2$ and a negligible small $K_V$ volume contribution, in line with previous reports[7,53]. These findings suggest that the roughness is not responsible for the two regimes behavior observed in the case of the Heusler samples.

Another possible physical mechanism which can explain the presence of the two regimes is the strain variation due to coherent–incoherent growth transition[54,55]. Within this model, below the critical thickness, the ferromagnetic layer grows uniformly strained in order to account for the lattice misfit with the adjacent layers. Above the critical thickness, the strains are partially relaxed through the formation of misfit dislocations. The changes in the magnetoelastic anisotropy contributions corresponding to this structural transition can be responsible for the presence of the two regimes[54,55]. This scenario is likely in the case of the Heusler samples, since both the bottom Pt layer the upper Heusler film grow out-of-plane textured. In order to test this hypothesis, we have deposited two additional sets of samples. The first set consisted of Si/SiO$_2$//Ta (6 nm)/ CFAS ($t_{CFAS}$)/MgO (1 nm)/Ta (3 nm) samples. The motivation to grow this type of samples was to obtain Heusler films with no out-of-plane texturing. Indeed, x-ray diffraction measurement [Fig. 1(b)], performed on a Ta/CFAS sample with a Heusler layer thickness of 10 nm, did not indicate the presence of any diffraction peaks, except for the one belonging to the Si substrate. This suggest that both the Ta and the CFAS films are either nanocrystalline or amorphous. Thus, in this type of structure we do not expect the presence of the coherent–incoherent growth transition. The second set of samples consisted of epitaxial MgO (001)//Cr (4 nm)/CFAS ($t_{CFAS}$)/MgO (1 nm)/Ta (3 nm) structures. The x-ray diffraction measurement [Fig. 1 (b)], performed on a Cr/CFAS sample with a Heusler layer thickness of 10 nm, indicates the exclusive presence of the (001) type reflections from the MgO substrate and the Cr and CFAS layers. This confirms the epitaxial growth of the stacks, except for the Ta capping layer, which is amorphous. Having in view the epitaxial growth we might expect for these samples a possible coherent–incoherent growth transition, eventually at higher CFAS thicknesses having in view the relative low mismatch between the CFAS lattice and the 45° in-plane rotated Cr lattice (0.7%). The effective magnetization dependence on the inverse thickness of the ferromagnetic layer for amorphous Ta/CFAS and epitaxial Cr/CFAS samples alongside with the Pt/CFAS samples is shown in Fig. 7. Is to be mentioned that in the for the Ta/CFAS samples the PMA was obtained for thicknesses below 1.6 nm, while for the Cr/CFAS the PMA was not achieved even for thicknesses down to 1 nm. Interestingly, in the case of the epitaxial Cr/CFAS samples, for which one might expect possible coherent–incoherent growth transition, a single linear behavior for the whole thickness range is observed. In the case amorphous Ta/CFAS samples, for which the coherent–incoherent growth transition is not expected, a two-regimes behavior can be distinguished. Although we cannot rule a possible coherent-incoherent growth transition at larger thicknesses, the results indicate this mechanism is not responsible for the two regimes behavior that we observe at relatively low thicknesses and other mechanisms must be at play.

By fitting the high thickness regime data from Fig.7 to eq. (5) we extracted for Ta/CFAS samples a surface anisotropy contribution $K_S$ of $0.27 \pm 0.08$ erg/cm$^2$. The volume contribution, $K_V$, was determined to the be negligible small, as expected for untextured films. Remarkably, the $K_S$ for the Ta/CFAS samples is similar within the error bars to one obtained for the Pt/CFAS samples. Moreover, even in the low thickness regime the $K_S$ might be assumed similar for the two sets of samples. However, we must consider the large uncertainty having in view the sparse data points available for fitting in the low thickness regime. Even so, the clear difference between the two sets of samples is that the Ta/CFAS one shows a larger critical thickness (around 2.4 nm) that separates the two anisotropy regimes, as compared to the Pt/CFAS one (around 1.5 nm). This suggests that the possible mechanism responsible for the two-regime behavior might be related to the atomic diffusion at the Pt/CFAS and Ta/CFAS interfaces. It is well known that Ta is prone to diffusion of light elements[56]. Therefore, a larger critical thickness for the Ta/CFAS samples will imply a larger atomic diffusion at the Ta/CFAS interface compared to the Pt/CFAS one.

To test the hypothesis of the interdiffusion, we performed Auger electron spectroscopy (AES) analyses on the three set of samples: Pt/CFAS, Cr/CFAS and Ta/CFAS. AES is a surface sensitive technique which can give information about the chemical composition of the surface with a depth detection limit of 1-2 nm. We started from 10 nm thick CFAS layer samples and first Ar ion etched the CFAS films down to 4 nm thickness and recorded the AES spectra. Subsequently, the Ar ion etching and AES spectra recording was repeated in steps of 1 nm until reaching the underlayer/CFAS interface. The etching rate of CFAS was previously calibrated using *ex-situ* x-ray reflectometry measurements. Figure 8(a) shows two spectra recorded for the Pt/CFAS sample, one after etching the CFAS layer down to 4 nm (Pt/CFAS 4 nm) and the other one after etching the CFAS layer down to 1 nm of thickness (Pt/CFAS 1 nm). In the case of the Pt/CFAS 4 nm spectrum the peaks of Co and Fe are visible alongside with the peaks from Al and Si. The inset of Fig. 8(a) depicts an enlargement of the Pt/CFAS 4 nm spectrum around the peaks of Al and Si. The amplitude of the Co and Fe peaks is much larger than the amplitude of the of Al and Si ones. This is due to the higher concentration and higher Auger relative sensitivity of the Co and Fe compared to the Al and Si. In the case of the Pt/CFAS 1 nm the spectrum shows the peaks from Co and Fe, with a lower amplitude, and the peaks from the Pt underlayer. The presence of the Co and Fe peaks together with the Pt peaks is not surprising. It is owed to the possible interdiffusion layer at the interface and to the finite depth resolution of the AES which probes both the CFAS layer and the Pt underlayer. The Al and Si peaks are not observable, which can be associated to the relatively low amplitude of the Al and Si falling below the detection limit of the measurement. To test this possibility, we acquired Auger spectra in a narrow energy window around the Al peak, using a longer acquisition time and averaging 10 spectra for each recoded spectrum. We selected the Al peak and not the Si one because of its larger amplitude. These spectra recorded for the Pt/CFAS, Ta/CFAS and Cr/CFAS samples after etching the CFAS layer down to 4, 3, 2 and 1 nm are shown in Fig.8 (b)-(d). In the case of the Pt/CFAS sample the Al peak is observable for CFAS thicknesses down to 1 nm, while in the case of Ta/CFAS for thicknesses down to 2 nm. Interestingly, in the case of the Cr/CFAS sample the Al peak is visible even for a CFAS thickness of 1 nm, although with lower amplitude. These findings suggest that at the underlayer/CFAS interface there is a diffusion of the lighter elements (Al and most likely also Si) towards the underlayer, with different degree, depending on the nature of the underlayer. As shown schematically in Fig. 8, due to this lighter elements diffusion a CoFe-rich layer forms at the underlayer/CFAS interface. The extent of the CoFe rich layer depends on the nature of the underlayer. It has the largest thickness for the Ta underlayer (between 2 and 3 nm), it is decreasing for the Pt underlayer (between 1 and 2 nm) and it is most likely non-existing or extremely thin (below 1 nm) in the case of the Cr underlayer.

The presence of the CoFe rich layer agrees with our findings concerning the occurrence of the high and the low effective PMA regimes depending on the thickness of the Heusler layer. In the case of the Pt/CFAS/MgO samples, the low effective PMA regime occurs for a CFAS layer thickness above 1.6 nm. In this case, the bottom interface consists of Pt/CoFe-rich layer, while the top one of CFAS/MgO. In principle, both interfaces could contribute to PMA through Co-O hybridization in the case of the Co-terminated CFAS/MgO interface[43] or through the d–d hybridization between the spin-split Co 3d bands and the Pt layer 5d bands with large spin-orbit coupling[44,45]. However, their contribution to PMA is small and, as we previously mentioned, would not stabilize perpendicular magnetization except for extremely thin CFAS layers. In the case of the high effective PMA regime (below 1.6 nm), the bottom interface is similar consisting of Pt/CoFe-rich layer and will contribute negligibly to PMA. However, the top interface is now constituted of CoFe-rich layer/MgO and will induce strong PMA through the hybridization of the [Co,Fe] $3d_{z^2}$ and O $2p_z$ orbitals[42]. The premise that the strong PMA is induced by the CoFe-rich layer/MgO interface is also consistent with our observations regarding the dependence of the magnetic anisotropy on the nature of the underlayer. As seen in Fig. 7, in the case of the Cr/CFAS samples, where no CoFe-rich layer was evidenced, there is only one anisotropy regime with a relatively low effective PMA. In the case of the Ta/CFAS samples, the high effective PMA regime is present starting from a larger CFAS thickness, as compared to de case of Pt/CFAS samples, which is in agreement with the thicker

CoFe-rich layer observed for the Ta/CFAS relative to the Pt/CFAS ones. It is to be mentioned that in the case of Ru/CFA/MgO and Cr/CFA/MgO annealed samples Al diffusion towards the MgO but not towards the underlayer was previously observed[46,47]. The lack of Al diffusion towards the Cr underlayer is in agreement with our findings. In the case of the aforementioned studies, the thermal annealing of the samples was necessary to facilitate the Al diffusion and to achieve strong PMA. In our case, for the Pt and Ta underlayer, we attain strong PMA in the low thickness regime without the need of thermal annealing. This indicates that for the Pt and Ta underlayers the [Al,Si] diffusion takes place during the growth of the CFAS film, which results in the formation of the interfacial CoFe-rich layer directly during deposition. A further deposition of MgO on this CoFe-rich layer will generate the strong PMA through the hybridization of the [Co,Fe] $3d_{z^2}$ and O $2p_z$ orbitals[42]. Having in view the similar behavior of the magnetic anisotropy for the CFA, CFAS, CFMS and CFS Heusler alloys thin films that we study here, it is reasonable to assume that in all the cases there is a diffusion of the lighter elements (Al, Si) towards the Pt underlayer and the formation of the CoFe-rich interfacial layer, which, when MgO is deposited on top, will give rise to the strong PMA in the low thickness regime.

We now discuss the thickness dependence of the Gilbert damping parameter extracted from the $\theta_H$ dependence of the linewidth $H_{PP}$. It is known that generally the linewidth is given by a sum of extrinsic and intrinsic contribution as[49,57-59]:

$$H_{\text{PP}} = H_{\text{PP}}^{\text{int}} + H_{\text{PP}}^{\text{ext}}, \tag{5}$$

$$H_{\text{PP}}^{\text{int}} = \alpha(H_1 + H_2)\left|\frac{dH_R}{d(\omega/\gamma)}\right|, \tag{6}$$

$$H_{\text{PP}}^{\text{ext}} = \left|\frac{dH_R}{d(4\pi M_{\text{eff}})}\right|\Delta(4\pi M_{\text{eff}}) + \left|\frac{dH_R}{d\theta_H}\right|\Delta\theta_H + \Delta H_{\text{TMS}}, \tag{7}$$

where, $\alpha$ is the intrinsic Gilbert damping parameter and the three terms in equation (7) are the linewidth enhancement due to the anisotropy distribution, due to deviation from planarity of the films and due to the two-magnon scattering. In the case of our films, the $\theta_H$ dependence of the linewidth $H_{PP}$ is well fitted using only the intrinsic contribution and the extrinsic enhancement due to the anisotropy distribution. For this, $|dH_R/d(\omega/\gamma)|$ and $|dH_R/d(4\pi M_{\text{eff}})|$ are numerically calculated using Eqs. (1)-(4) and the $H_{PP}$ vs. $\theta_H$ experimental dependence is fitted to Eq. (5) using $\alpha$ and $\Delta(4\pi M_{\text{eff}})$ as adjustable parameters[49]. An example of a fit curve is depicted in Fig. 4(b) for the case of the 2.4 nm thick Pt/CFAS sample. Figure 9 shows the $\alpha$ dependence on the inverse ferromagnetic layer (1/t) thickness for the Pt/CFA, Pt/CFS, Pt/CFMS, Pt/CFAS and Pt/CFB samples. We will first discuss the case of CFB, where a linear dependence is observed. The linear increase of the Gilbert damping parameter with 1/t is expected and it is due to the angular momentum loss due to the spin pumping effect in the Pt layer. In this type of structures it was shown[60] that the total damping is given by $\alpha = \alpha_0 + \alpha_{SP}/t$, where $\alpha_0$ is the Gilbert damping of the ferromagnetic film and $\alpha_{SP}$ is due to the spin pumping effect. By linear fitting the data in Fig. 9 we obtain a Gilbert damping parameter for the CFB of 0.0028 ± 0.0003, in agreement with other reports[61,62]. In the case of the CFAS films, the linear dependence is observed only for the large thickness region and by fitting this data we obtain a Gilbert damping parameter of 0.0053 ± 0.0012, consistent with previously reported values for relatively thicker films[25]. The low thickness data deviates from the linear dependence. This behavior is similar for all the other studied Heusler films, with the low thickness deviation being even more pronounced. The strong increase of the damping can be related to the [Al,Si] diffusion and the formation of the interfacial CoFe-rich layer. Since the [Al,Si] diffusion is more important for thinner films, it will have a stronger impact on the chemical composition relative to the thicker ones. The relatively small damping of the Co based full Heusler alloys is a consequence of their specific electronic structure[21]. Consequently, deviations from the correct stoichiometry, which is expected to have an important effect on the electronic structure, will lead to a strong increase of the damping, as shown, for example, by *ab-initio* calculation in the case of Al deficient CFA films[47]. Therefore, the increase of the damping beyond the spin pumping effect for the thinner Heusler films is explained by the interfacial CoFe-rich layer formation.

*Conclusions*

We have studied the mechanisms responsible for PMA in the case of $Co_2FeAl$, $Co_2FeAl_{0.5}Si_{0.5}$, $Co_2FeSi$ and $Co_2Fe_{0.5}Mn_{0.5}Si$ Heusler alloy thin films sandwiched between Pt and MgO layers. We showed that the ultrathin Heusler films exhibit strong PMA irrespective of their composition. The effective magnetization displays a two-regime behavior depending on the thickness of the Heusler layers. The two-regime behavior is generated by the formation of an CoFe-rich layer at the underlayer/Heusler interface due to the interdiffusion. The strong PMA observed in the case of the ultrathin films can be explained by the electronic hybridization of the CoFe-rich metallic layer and oxygen orbitals across the ferromagnet/MgO interface. The formation of the interfacial CoFe-rich layer causes the increase of the Gilbert damping coefficient beyond the spin pumping for the ultrathin Heusler films. Our results illustrate that the strong PMA is not an intrinsic property of the Heusler/MgO interface, but it is actively influenced by the interdiffusion, which can be tuned by a proper choice of the underlayer material.

FIG. 1. (a) *2θ/ω* x-ray diffraction patterns recorded for four representative Pt/Co$_2$YZ/MgO samples having a thickness of the Heusler layer of 10 nm. The patterns show the (111) and (222) peaks belonging to the Pt layer, the (022) peak from the Heusler films and the (001) peak of the Si substrate. (b) *2θ/ω* x-ray diffraction patterns for the Ta/CFAS (10 nm)/MgO and Cr/CFAS (10 nm)/MgO samples indicating the amorphous or epitaxial growth of the CFAS layer, respectively.

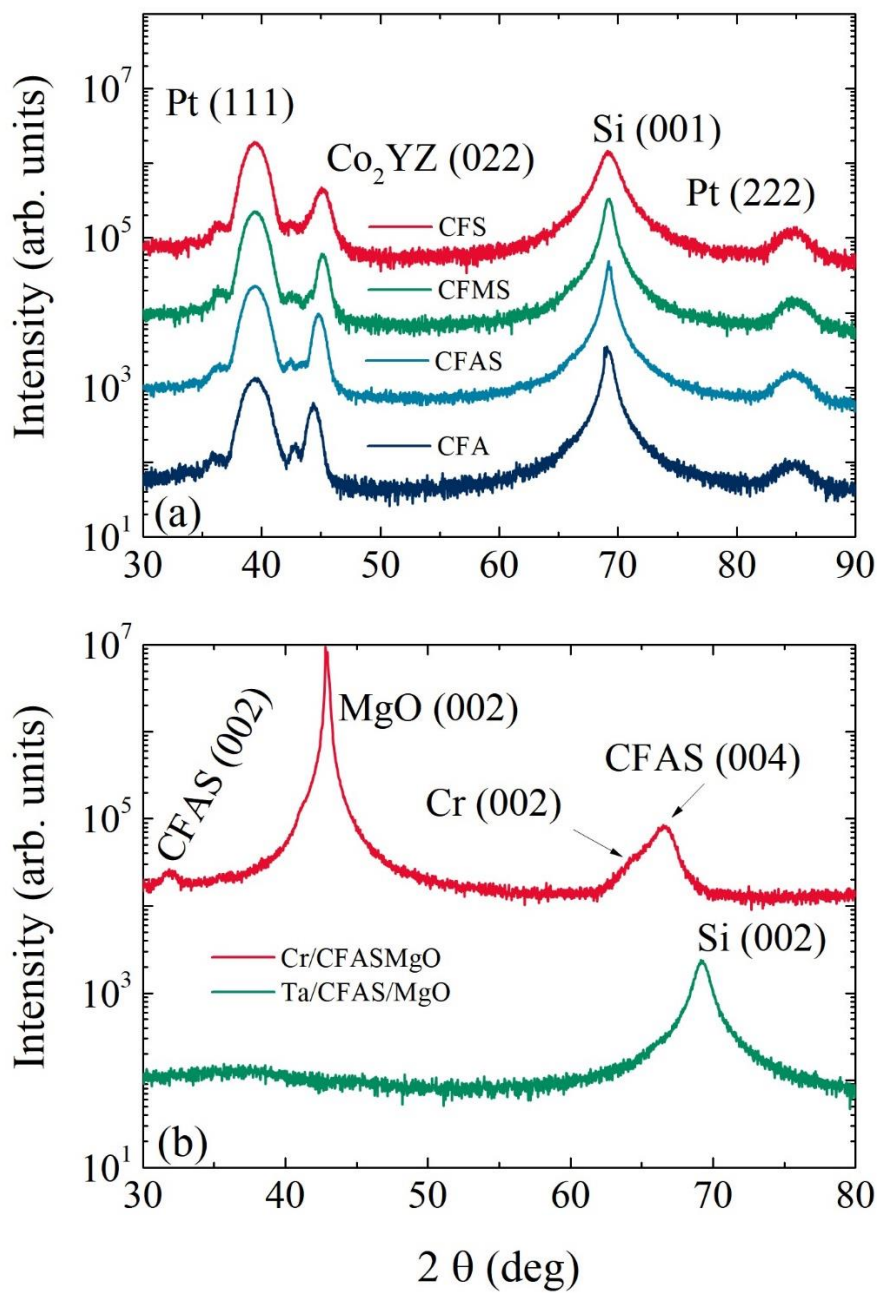

FIG. 2. Hysteresis loops measured with the magnetic field applied perpendicular to the plane of the samples. Depending on the thickness of the Heusler layers, the samples show in-plane magnetic anisotropy (a)-(d) or perpendicular magnetic anisotropy (e)-(h).

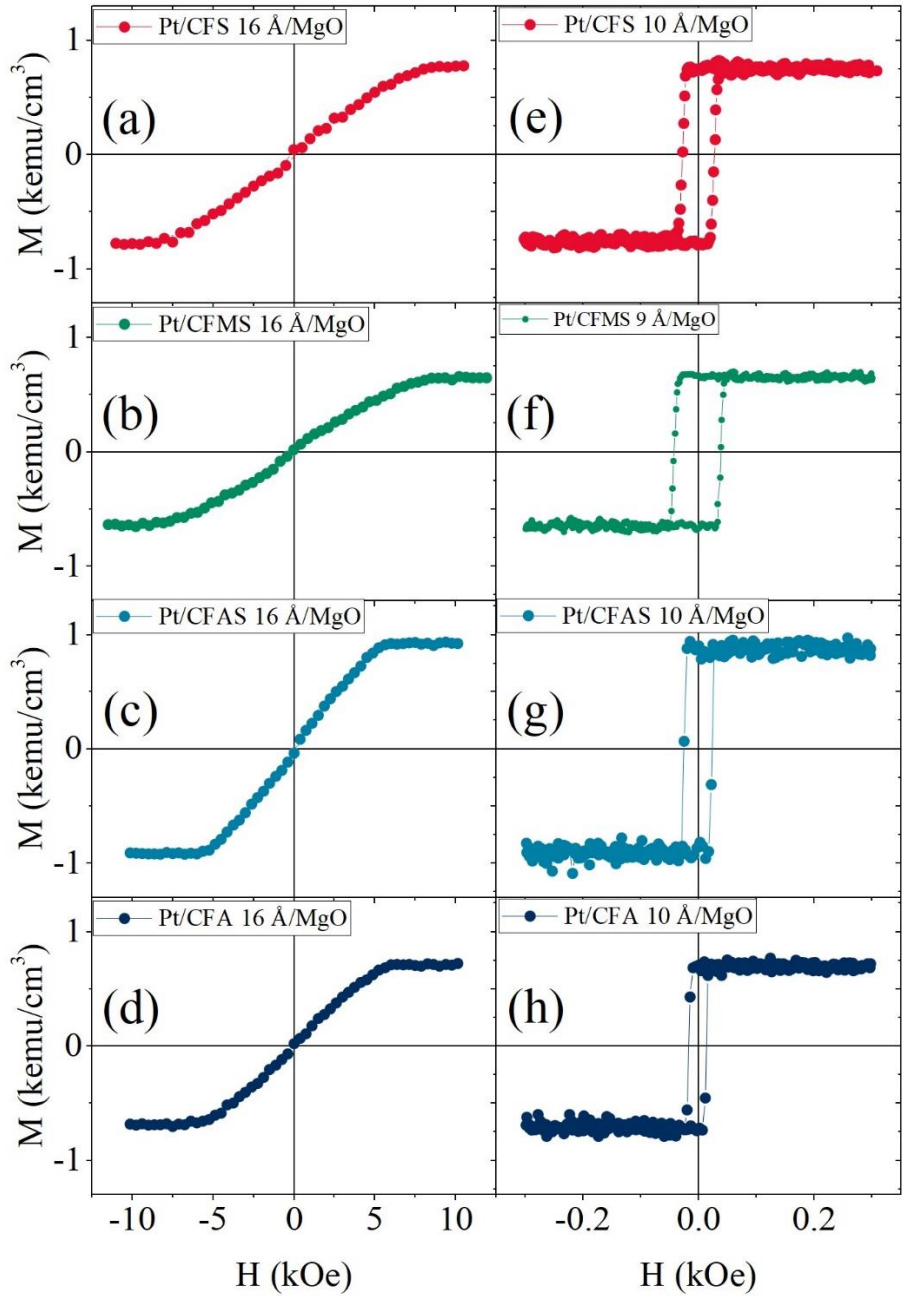

FIG. 3. Typical FMR spectra measured at 9.79 GHz for different $\theta_H$ field angles for a 2.4 nm thick Pt/CFAS sample.

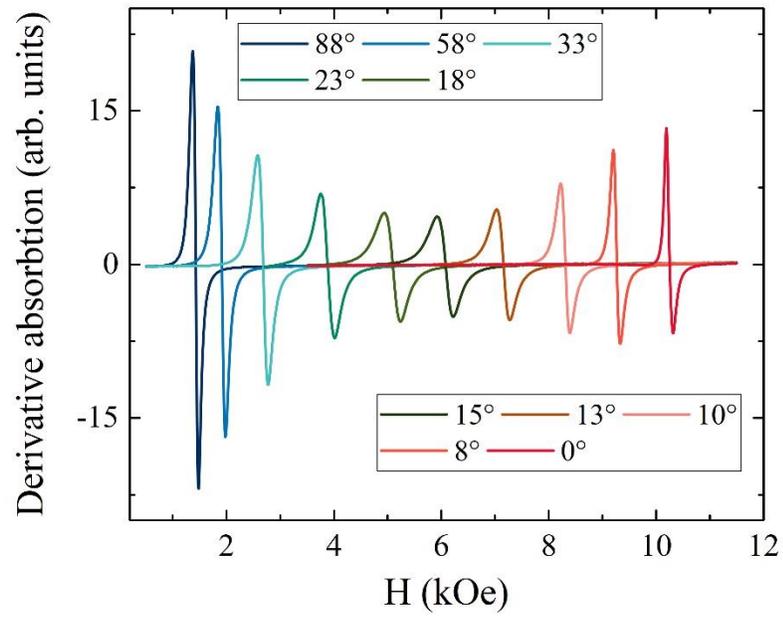

FIG. 4. (a) Resonance field $H_R$ and (b) linewidth $H_{PP}$ dependence on the $\theta_H$ field angle for a 2.4 nm thick Pt/CFAS sample. The inset shows a schematic of the measurement geometry. The points stand for experimental data while the lines represent the result of the theoretical fits, as described in text.

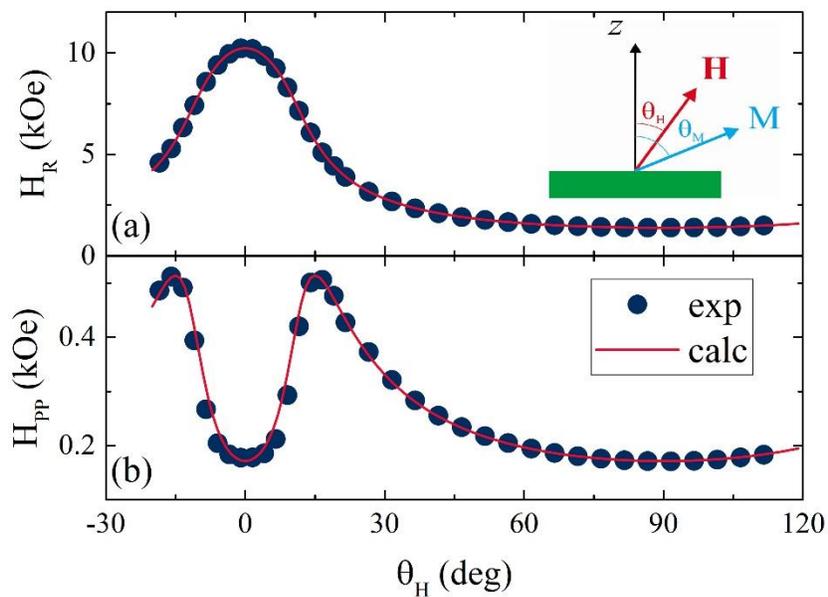

FIG. 5. *g* factor dependence on the thickness of the Heusler layers for samples with different Heusler layer composition.

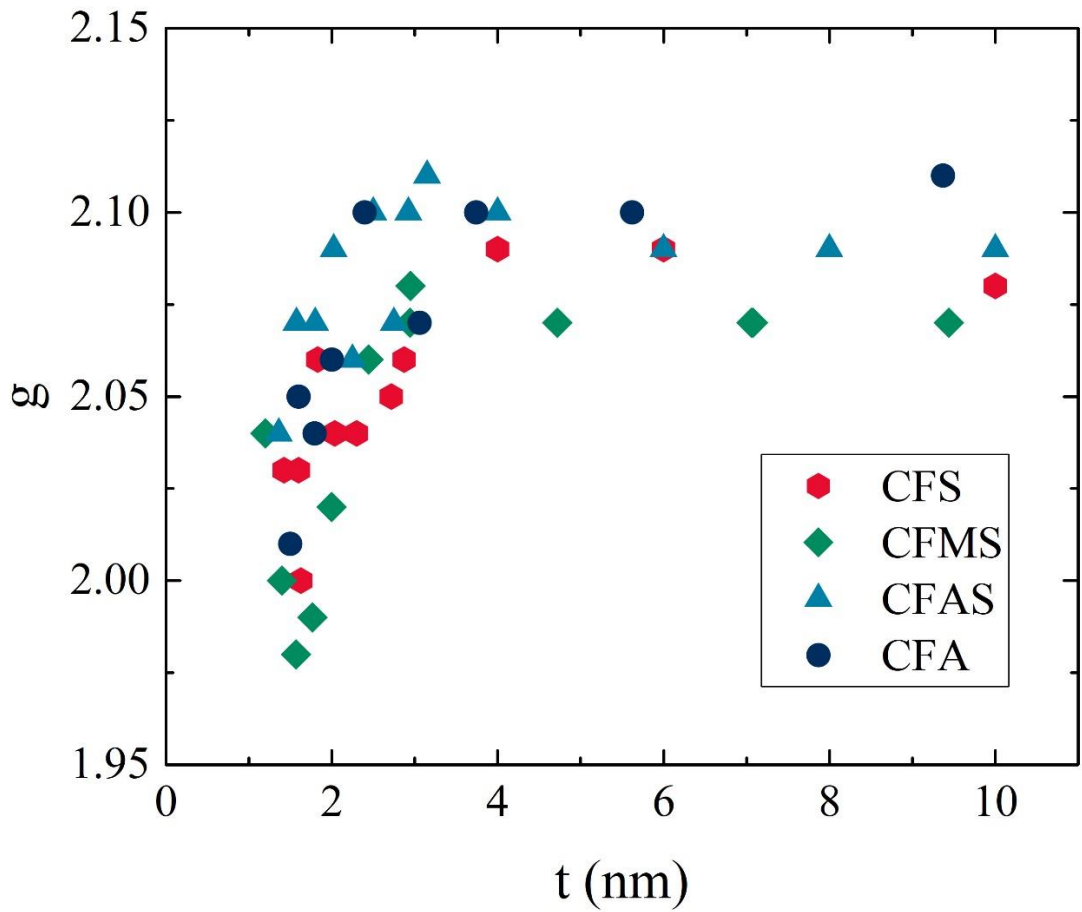

FIG. 6. The effective magnetization $4\pi M_{eff}$ dependence on the inverse thickness of the ferromagnetic layer for samples with different compositions. The points are experimental data while the lines are linear fits. In the case of the Heusler samples two linear fits correspond to the two anisotropy regimes.

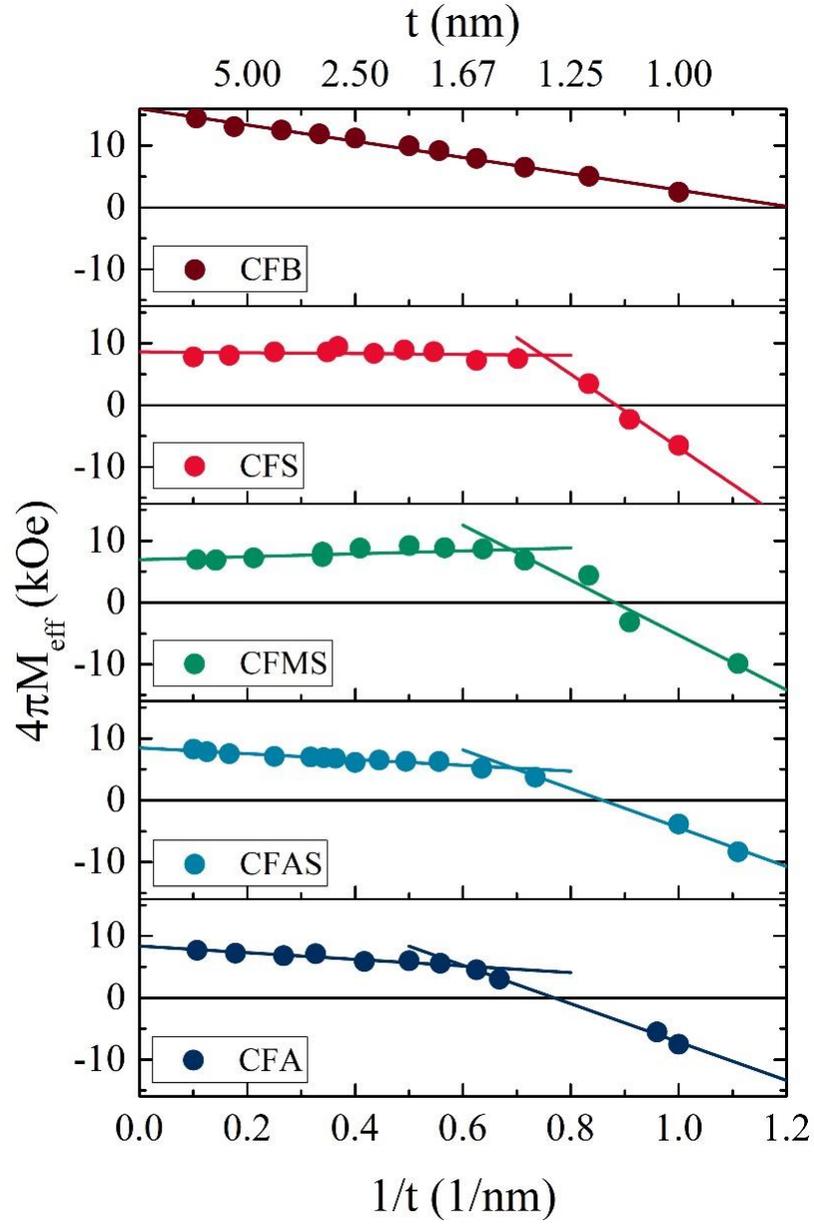

FIG. 7. The effective magnetization $4\pi M_{eff}$ dependence on the inverse thickness of the ferromagnetic layer for amorphous Ta/CFAS and epitaxial Cr/CFAS samples. The data for Pt/CFAS is also shown for comparison. The points are experimental data while the lines are linear fits.

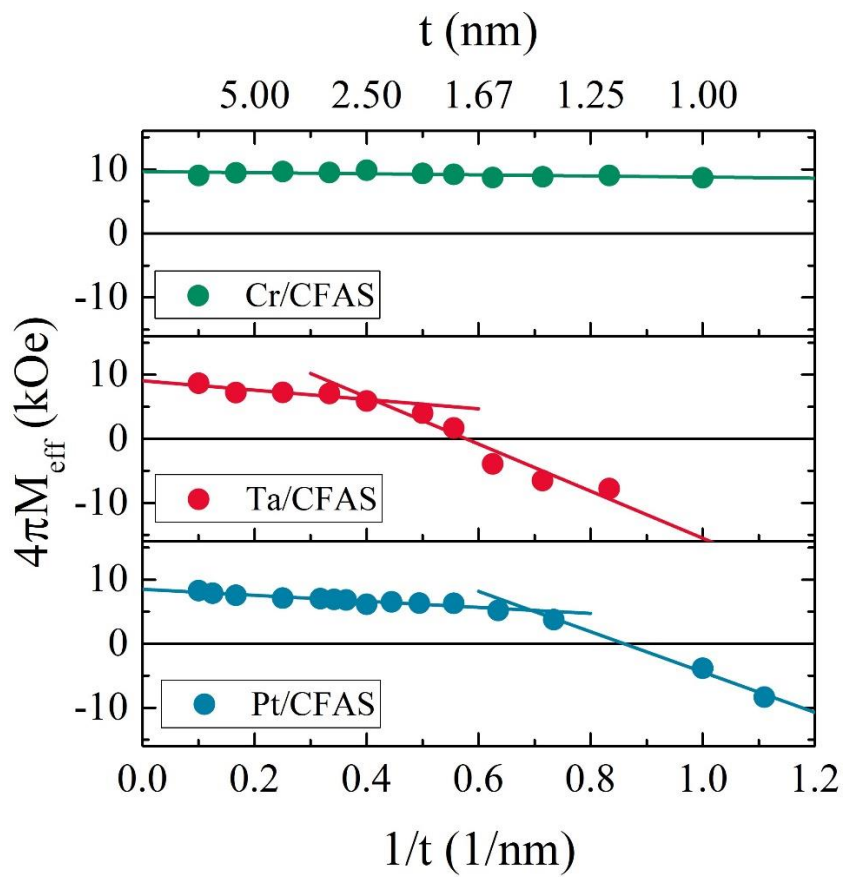

FIG. 8. (a) AES spectra recoded for the Pt/CFAS sample after etching the CFAS layer down to 4 and 1 nm, respectively. The inset shows a zoom around de Al and Si peaks. AES spectra recorded around the Al peak after etching the CFAS layer down to 4, 3, 2 and 1 nm for the (b) Pt/CFAS, (c) Ta/CFAS and (d) Cr/CFAS samples. Schematic representation of the [Al,Si] diffusion towards the underlayer and the interfacial CoFe-rich layer formation.

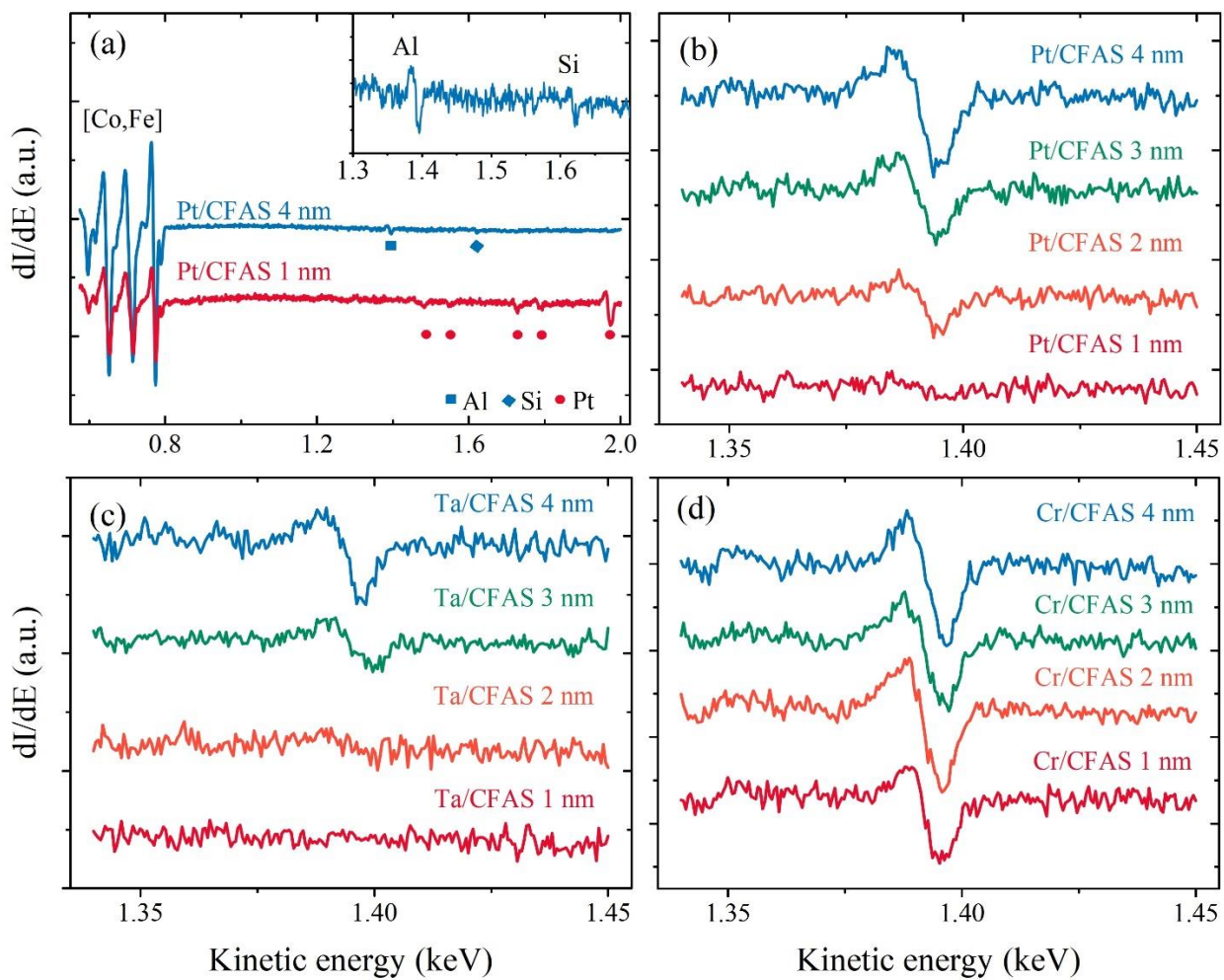

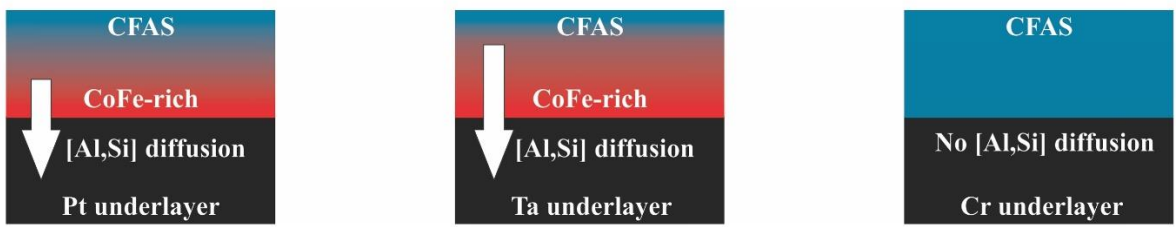
19

FIG. 9. Gilbert damping parameter ($\alpha$) dependence on the inverse ferromagnetic layer (1/t) thickness for the Pt/CFA, Pt/CFAS, Pt/CFMS, Pt/CFS and Pt/CFB samples. The points are experimental data while the lines are linear fits for Pt/CFB and Pt/CFAS samples. In the case of the Pt/CFAS samples only the linear large thickness range was used for fitting.

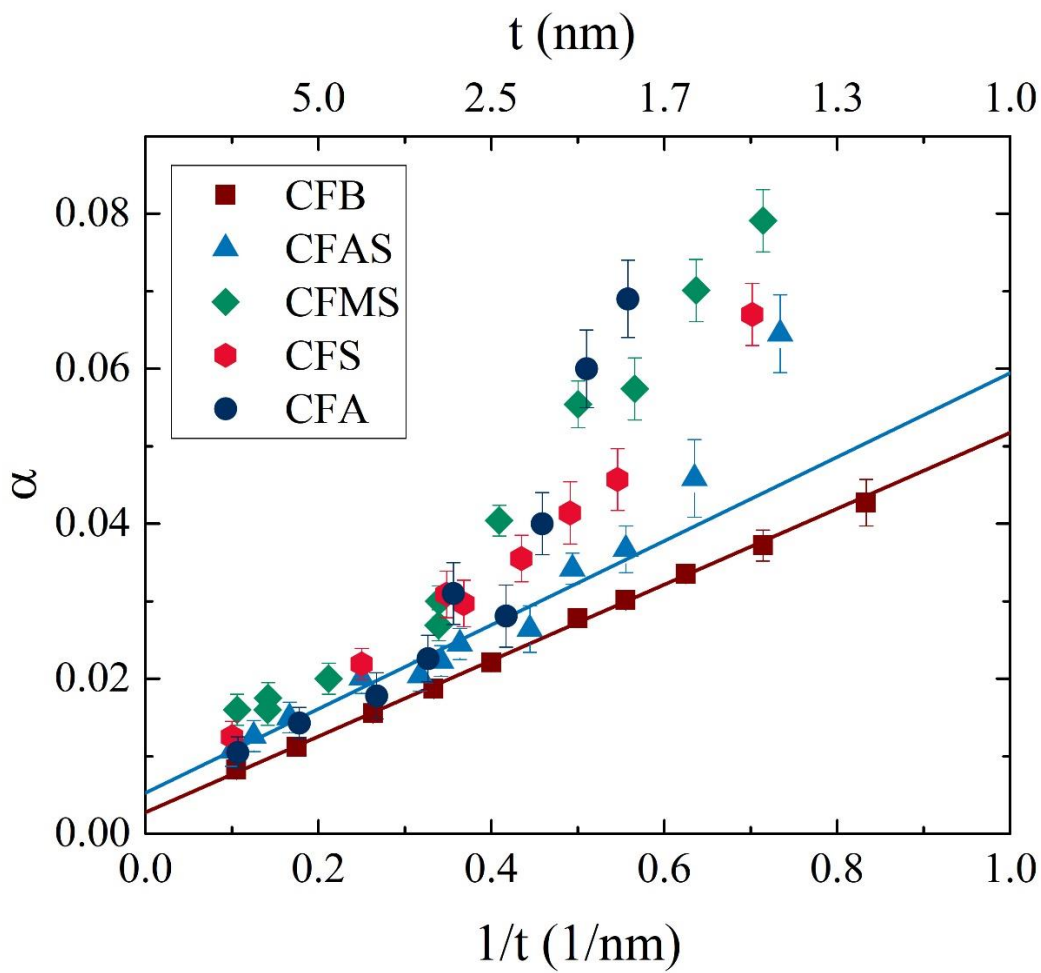